\documentclass[amsmath,amssymb,aps,showpacs,twocolumn,superscriptaddress,letterpaper,longbibliography,floatfix]{revtex4-2}

\usepackage{graphicx}
\usepackage{dcolumn}
\usepackage{bm}
\usepackage{hyperref}
\usepackage{lineno}
\usepackage{natbib} 
\usepackage{amsmath, amssymb} 
\usepackage{color}
\usepackage[dvipsnames]{xcolor}
\usepackage{comment} 
\usepackage{hyperref} 
\usepackage{float} 
\usepackage{makecell} 
\usepackage{cleveref}
\usepackage{tikz}
\usepackage{gensymb}
\usepackage[table,xcdraw]{xcolor}

\hypersetup{breaklinks = true, colorlinks = true, citecolor = blue, linkcolor = blue, urlcolor = blue}

\begin{document}
\title{Event Generator Tuning as a Robustness Test}
\author{Jean Wolfs}
\email{jwolfs@ur.rochester.edu}
\affiliation{Department of Physics and Astronomy, University of Rochester}
\author{Chris M. Marshall}
\email{chris.marshall@rochester.edu}
\affiliation{Department of Physics and Astronomy, University of Rochester}
\date{\today}


\begin{abstract}

Neutrino oscillation experiments use Monte Carlo event generators to predict neutrino-nucleus interactions. Cross section uncertainties are typically implemented by varying the parameters of the model(s) used in the generator. We study the performance of two commonly-used model configurations of the GENIE generator (G18\_10a\_02\_11a and AR23\_0i\_00\_000) and their uncertainties by tuning parameters to cross section data, and then comparing the resulting tuned prediction to a suite of other measurements from T2K, MicroBooNE, and MINERvA. This reveals whether the model can simultaneously describe several datasets, as well as whether the uncertainties are adequately robust. We find that G18 and especially AR23 are reasonable in predicting lower-energy measurements from T2K and MicroBooNE, but unable to describe MINERvA data, and discuss the implications for short-baseline oscillation searches. We attempt to replicate a tuning procedure developed by MicroBooNE using several different measurements, and find substantially different results depending on which measurement is used, and that the MicroBooNE tune does not agree with other measurements. We conclude that the SBN experiment should \textit{not} tune its generator to external data.

\end{abstract}

\maketitle


\section{Introduction}\label{sec:intro}

The discovery of neutrino oscillations~\cite{SK,SNO} implies that neutrinos have mass and that there is non-trivial mixing between the neutrino flavor states and the neutrino mass states. In an approximation with only two neutrino flavors, $\alpha$ and $\beta$, and two neutrino masses, $m_1$ and $m_2$, the vacuum probability of neutrino oscillations is:
\begin{equation}\label{eqn:prob}
    P_{\alpha \rightarrow \beta}(E_\nu)=\sin^2(2\theta)\sin^2\left(\frac{\Delta m^2 L}{4E_\nu}\right)
\end{equation}
where $\theta$ is the mixing angle, $\Delta m^2=(m_2^2-m_1^2)$ is the mass splitting, $L$ is the baseline, and $E_\nu$ is the neutrino energy. 

With three flavors ($\nu_e$, $\nu_\mu$, $\nu_\tau$), there are three mixing angles ($\theta_{12}$, $\theta_{23}$ and $\theta_{13}$) and three mass splittings ($\Delta m^2_{21}$, $\Delta m^2_{32}$ and $\Delta m^2_{31}$), only two of which are independent because $\Delta m^2_{32}=\Delta m^2_{31}-\Delta m^2_{21}$. Given the measurements of $\Delta m^2_{21}=(7.53 \pm 0.18) \times 10^{-5} \text{ eV}^2$ and $\Delta m^2_{32}=(2.455 \pm 0.028) \times 10^{-3} \text{ eV}^2$~\cite{PDG}, Eqn.~\ref{eqn:prob} is maximized and neutrino oscillations are most probable when $L/E_\nu \simeq 16,000 \text{ km/GeV}$ and $L/E_\nu \simeq 500 \text{ km/GeV}$.

Accelerator-based neutrino oscillation experiments measure Eqn.~\ref{eqn:prob} indirectly by measuring the interaction rate for some neutrino flavor as a function of reconstructed neutrino energy:
\begin{equation}\label{eqn:rate}
    \frac{dN_\beta}{dE_{rec}} = \int \Phi_{\alpha}(E_\nu)P_{\alpha \rightarrow \beta}(E_\nu)\sigma_\beta(E_\nu)T_\beta(E_\nu,E_{rec})dE_\nu
\end{equation}
where $E_{rec}$ is the reconstructed neutrino energy, $\Phi_{\alpha}$ is the unoscillated neutrino flux, $P_{\alpha \rightarrow \beta}$ is the oscillation probability from Eqn.~\ref{eqn:prob}, $\sigma_\beta$ is the cross section for $\nu_\beta$, and $T_\beta$ is the detector response. A measurement of Eqn.~\ref{eqn:rate} is called an appearance measurement when $\alpha\neq\beta$ and a disappearance measurement when $\alpha=\beta$.

Of the variables on the right hand side of Eqn.~\ref{eqn:rate}, $E_{rec}$ is measured in the experiment, while $\Phi_\alpha$, $\sigma_\beta$, and $T_\beta$ are estimated using models, often informed by external data. The neutrino flux $\Phi_\alpha$ is typically estimated using a simulation of the neutrino beam. In the case of the Booster Neutrino Beam (BNB) experiments at Fermilab, a Geant4-based software package is used to predict the flux~\cite{geant4,bnb_flux}. The systematic uncertainty in the BNB flux estimation is dominated by uncertainties in the rate and energy of secondary mesons produced in proton interactions with the target, although uncertainties in beam focusing are also relevant.

The second estimated variable in Eqn.~\ref{eqn:rate}, the cross section $\sigma_\beta$, is calculated by a neutrino event generator. Modern experiments use GENIE~\cite{genie} or NEUT~\cite{neut1,neut2,neut3} in their primary analyses, but may use GiBUU~\cite{gibuu} or NuWro~\cite{nuwro} in other studies. Each generator uses a series of models to simulate the neutrino interaction itself, as well as the kinematics of the initial nuclear state and final state interactions (FSI) between outgoing particles and the residual nucleus. Finally, the detector response function, $T_\beta$, relates true neutrino energy to a detector observable, typically reconstructed neutrino energy. In addition to detector effects, $T_\beta$ is sensitive to exclusive neutrino cross section estimates because the detector response can vary for different final states.

Since the mid-1990s, several experiments have made measurements of Eqn.~\ref{eqn:prob} that are inconsistent with three-flavor predictions. The LSND experiment measured a $3.8\sigma$ excess of $\bar{\nu}_e$ events in a reactor-based $\bar{\nu}_\mu$ beam~\cite{LSND}. Then, MiniBooNE measured a $4.5\sigma$ excess of $\nu_e$ and a $2.8\sigma$ excess of $\bar{\nu}_e$ events in accelerator-based $\nu_\mu$ and $\bar{\nu}_\mu$ beams, respectively. Together, LSND and MiniBooNE's measurements are known as the ``short baseline anomaly". Meanwhile, the GALLEX and SAGE experiments measured the rate of neutrino capture on gallium using the radioactive sources $^{51}$Cr and $^{37}$Ar and found a combined deficit of $3\sigma$ $\nu_e$ events~\cite{sage_cr51,sage_ar37}. This result is known as the ``gallium anomaly.''

The short baseline and gallium anomalies could be explained by the existence of a fourth neutrino mass state with mass splittings $\Delta m^2_{43}\approx\Delta m^2_{42}\approx\Delta m^2_{41}\sim \mathcal{O}(1\text{ eV}^2)$. As measurements of the $Z$ boson width strongly disfavor the existence of more than three weakly interacting neutrinos~\cite{z_resonance}, this fourth neutrino would be a ``sterile neutrino'' that could only interact via the gravitational force.

Experiments like LSND and MiniBooNE that use accelerator-based neutrino beams can be categorized by the ratio of their baseline to their energy, $L/E_\nu$ in Eqn.~\ref{eqn:prob}. Long-baseline (LBL) experiments have $L/E_\nu$ close to the oscillation maximum at $L/E_\nu \simeq500 \text{ km/GeV}$, and so are designed to measure standard oscillations due to mixing with the third mass state. On the other hand, short baseline (SBL) experiments have $L/E_\nu$ close to the oscillation minimum at $L/E_\nu \simeq 1 \text{ km/GeV}$, and so are designed to search for non-standard oscillations mediated by heavy sterile neutrinos.

LBL experiments typically operate two detectors: a near detector (ND) close to the neutrino source to measure the initial, un-oscillated beam and a far detector (FD) at a distance $L$ from the neutrino source to measure oscillations.  Most SBL experiments are conducted with only a single detector because it is difficult to position a detector close enough to the neutrino source that rapid, sterile oscillations would be truly negligible. As a consequence, single-detector SBL experiments must estimate $\Phi_\alpha$, $\sigma_\beta$, and $T_\beta$ without the aid of ND constraints, leading to larger systematic uncertainties in their measurements of Eqn.~\ref{eqn:rate}.

\begin{figure}
    \centering
    \includegraphics[width=\linewidth]{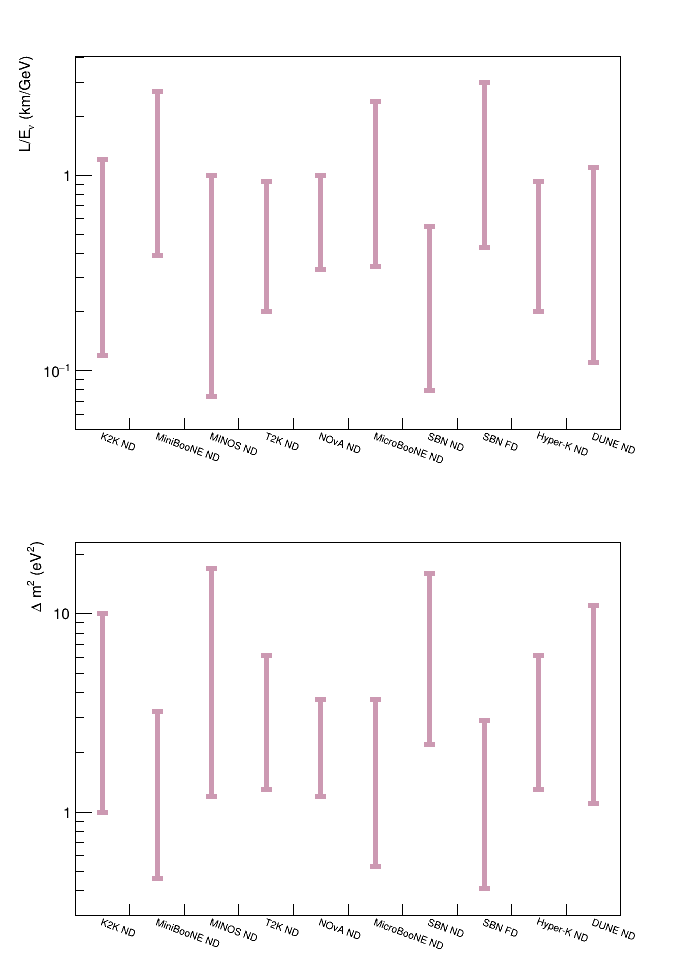}
    \caption{The sensitivity ranges of the accelerator-based neutrino experiments from Table~\ref{tab:experiments} to $L/E_\nu$ (left) and $\Delta m^2$ (right). The LBL far detectors from Table~\ref{tab:experiments} are excluded, as they all have sensitivity in the few $\times 10^{-3}$ eV$^2$ region. }
    \label{fig:experiments}
\end{figure}

Table~\ref{tab:experiments} summaries a number of past, present, and future accelerator-based neutrino experiments by listing the approximate range of $L/E_\nu$ and $\Delta m^2$ to which they are sensitive in both the ND and the FD. The energy ranges in the table correspond roughly to where the event rate is at least 10\% of the peak. The baseline is taken to be the distance to the production target; for near detectors this is a rough approximation for $L/E_\nu$ because line source effects are not negligible. Fig.~\ref{fig:experiments} plots the ranges of $L/E_\nu$ and $\Delta m^2$ from Table~\ref{tab:experiments} for those detectors with $\Delta m^2>10^{-2}\text{ eV}^2$. The upper range of $\Delta m^2$ corresponds to where the experiment could observe spectral distortions due to a sterile neutrino; experiments could also observe averaged effects due to rapid oscillations at higher $\Delta m^2$.

\begin{table*}[]
\centering
{\renewcommand{\arraystretch}{1.2}
\begin{tabular}{|c|c|c|c|c|c|c|c|c|c|}
\hline
                    & \textbf{First Data} & \textbf{\begin{tabular}[c]{@{}c@{}}FD\\ Target\end{tabular}} & \textbf{\begin{tabular}[c]{@{}c@{}}\boldmath$E_\nu$\\ (GeV)\end{tabular}} & \textbf{\begin{tabular}[c]{@{}c@{}}\boldmath$L_{ND}$\\ (km)\end{tabular}} & \textbf{\begin{tabular}[c]{@{}c@{}}\boldmath$L_{ND}/E_\nu$\\ (km/GeV)\end{tabular}} & \textbf{\begin{tabular}[c]{@{}c@{}}\boldmath$\Delta m^2_{ND}$\\ (eV\boldmath$^2$)\end{tabular}} & \textbf{\begin{tabular}[c]{@{}c@{}}\boldmath$L_{FD}$\\ (km)\end{tabular}} & \textbf{\begin{tabular}[c]{@{}c@{}}\boldmath$L_{FD}/E_\nu$\\ (km/GeV)\end{tabular}} & \textbf{\begin{tabular}[c]{@{}c@{}}\boldmath$\Delta m^2_{FD}$\\ (eV\boldmath$^2$)\end{tabular}} \\ \hline
\textbf{K2K}~\cite{k2k_general}        & 1999 & H$_2$O               & 0.3-2.5                                                                  & 0.3                                                                       & 0.12-1.2                                                                            & 1.0-10                                                                                          & 250                                                                       & 100-1000                                                                            & $(1.2-10) \times 10^{-3}$                \\ \hline
\textbf{MiniBooNE}~\cite{miniboone_general}  & 2002 & CH$_2$               & 0.2-1.4                                                                   & 0.54                                                                       & 0.4-2.7                                                                                 & 0.5-3.2                                                                                             & N/A                                                                      & N/A                                                                            & N/A                                                                                        \\ \hline
\textbf{MINOS}~\cite{minos_general}      & 2005 & Fe               & 1-14                                                                      & 1.04                                                                      & 0.07-1.0                                                                           & 1.2-17                                                                                          & 735                                                                       & 53-735                                                                              & $(1.7-20) \times 10^{-3}$                          \\ \hline
\textbf{T2K}~\cite{t2k_general}        & 2010 & H$_2$O               & 0.3-1.4                                                                   & 0.28                                                                      & 0.20-0.93                                                                           & 1.3-6.2                                                                                         & 295                                                                       & 211-983                                                                             & $(1.3-5.9) \times 10^{-3}$                                                                                   \\ \hline
\textbf{NOvA}~\cite{nova_general}       & 2014 & C,Cl,H               & 1.0-3.0                                                                       & 1.0                                                                         & 0.33-1.0                                                                            & 1.2-3.7                                                                                         & 810                                                                       & 270-810                                                                             & $(1.5-4.6) \times 10^{-3}$                                                                                   \\ \hline
\textbf{MicroBooNE}~\cite{microboone_general} & 2015 & Ar               & 0.2-1.4                                                                   & 0.47                                                                       & 0.3-2.4                                                                                 & 0.5-3.7                                                                                             & N/A                                                                      & N/A                                                                            & N/A                                                                                        \\ \hline
\textbf{SBN}~\cite{sbn_general}        & 2021 & Ar               & 0.2-1.4                                                                   & 0.11                                                                      & 0.079-0.55                                                                          & 2.2-16                                                                                          & 0.6                                                                       & 0.43-3.0                                                                            & 0.41-2.9                                                                                        \\ \hline
\textbf{Hyper-K}~\cite{hyperk_general}    & 2027 & H$_2$O               & 0.3-1.4                                                                   & 0.28                                                                      & 0.20-0.93                                                                           & 1.3-6.2                                                                                         & 295                                                                       & 211-983                                                                             & $(1.3-5.9) \times 10^{-3}$                                                                                   \\ \hline
\textbf{DUNE}~\cite{dune_general}       & 2030 & Ar               & 0.5-5.0                                                                     & 0.57                                                                     & 0.11-1.1                                                                            & 1.1-11                                                                                          & 1285                                                                      & 257-2570                                                                            & $(4.8-48) \times 10^{-3}$                                                                                   \\ \hline
\end{tabular}
}
\caption{A (non-comprehensive) list of LBL and SBL accelerator-based neutrino experiments and the ranges of $L/E_\nu$ and $\Delta m^2$ that they probe. For $\Delta m^2$ within the range quoted, oscillations due to that mass splitting would produce a spectral effect as a function of neutrino energy within the detector. For $\Delta m^2$ above the quoted upper bound, the detector would see an averaged oscillation effect across its neutrino energy spectrum.}
\label{tab:experiments}
\end{table*}

Two of the experiments from Table~\ref{tab:experiments} were designed to test the short baseline anomaly. MicroBooNE was an 85-ton liquid argon time projection chamber (LArTPC) positioned at $L=470$~m from the BNB target that operated from 2015 to 2021. MicroBooNE searched for $\nu_e$ appearance, but found no excess of events like LSND and MiniBooNE~\cite{uboone_lee_nue,uboone_lee_nue_inc,uboone_lee_nue_0pi,uboone_lee_nue_ccqe,uboone_lee_photon}. The follow-up Short-Baseline Neutrino (SBN) experiment will further test the short baseline anomaly~\cite{sbn_proposal}. The SBN ND, called the Short-Baseline Near Detector (SBND), is a 112-ton LArTPC positioned at $L=110$~m from the BNB target and the SBN FD, ICARUS, is a 476-ton LArTPC positioned at $L=600$~m from the BNB target. SBN is the first two-detector SBL experiment.

The interaction cross section $\sigma_\beta$ is a critical input to MicroBooNE and SBN's oscillation searches. Its estimation using an event generator such as GENIE depends on numerous uncertain parameters, giving the estimation an \textit{a priori} uncertainty at the 10-20\% level. In two-detector experiments, this uncertainty can be reduced by fitting to the ND, where oscillations are negligible~\cite{t2k_oscil_par_measurement,nova_systematic_uncertainties}).

Another plausible way to reduce the cross section uncertainty is to tune the parameters of the event generator to external cross section data. The GENIE collaboration fits its own interaction models to data collected over a wide phase space and makes these tunes publicly available~\cite{genie_tune}. Some experiments use these official GENIE tunes in their analyses, while others fit the interaction model themselves to data from their phase subspace. MicroBooNE, for example, tuned its GENIE model to a T2K cross section that was made using a similar neutrino flux as MicroBooNE~\cite{ub_tune,t2k_cross_2016}.

We expand on the approach of MicroBooNE~\cite{ub_tune} by tuning two commonly used model configurations to five different cross section measurements, and evaluate the performance by comparing the tuned model predictions to \textit{other} data. This provides a way to test the robustness of the model and its uncertainty parameterization. If a model has the correct physics, one would expect tunes of that model to describe all of the available data with the same parameter values. If an interaction model is incomplete but its uncertainties are robust, one would expect tunes of that model to reasonably describe available data, but perhaps with different best-fit parameter values. We also attempt to apply the specific procedure used by MicroBooNE to other measurements, in order to study how the model used in MicroBooNE is dependent on which specific measurement is tuned to. The study details are give in Section~\ref{sec:tuning}, and results are presented in Section~\ref{sec:results}.

\section{Analysis method}\label{sec:tuning}

Two commonly-used configurations of the GENIE version 3.4.0 interaction model are fit to a variety of cross section measurements. The G18\_10a\_02\_11a configuration (henceforth referred to as G18) is used by several neutrino experiments, including MicroBooNE~\cite{genie_tune,ub_tune}. The AR23\_0i\_00\_000 configuration (henceforth AR23) was developed by the DUNE collaboration for use in its oscillation sensitivity analyses, and is also used by SBN. The physics models used in these two comprehensive model configurations (CMCs) are outlined in Table~\ref{tab:genie_tunes}.

The two configurations use the same resonance, inelastic, hadronization, and FSI models, but different ground state models. AR23 uses a modified Local Fermi Gas (LFG) model which populates a broader region of missing energy and momentum space. Both G18 and AR23 use the Valencia charged-current quasi-elastic (CCQE) cross section model~\cite{valencia_ccqe,valencia_ccqe_erratum}, although G18 uses the historical dipole approximation for the axial form factor, while AR23 uses a $z$-expansion, which is expected to be more realistic at high $Q^{2}$~\cite{zexpansion}. G18 also uses the Valencia model for the charged current 2-particle, 2-hole (2p2h) cross section~\cite{valencia_2p2h}, which only predicts three-momentum transfer below 1.2 GeV; AR23 uses the fully-relativistic SuSAv2 2p2h model~\cite{susav2_2p2h}, which does not have a three-momentum transfer ceiling.

Alternative neutrino event generators like NEUT~\cite{neut1,neut2,neut3} and NuWro~\cite{nuwro} also support multiple model configurations with physics models that are not dissimilar to the ones in G18 and AR23. For example, the NEUT configuration chosen by T2K for its oscillation analyses~\cite{t2k_oscil_par_measurement} uses the ``Benhar spectral function'' ground state model~\cite{spectral_fcn} that is qualitatively similar to AR23's, the same 2p2h model as G18, and a resonance model that is the precursor to the one in both G18 and AR23. The GiBUU~\cite{gibuu} is unlike the others in thatit relies on transport theory~\cite{gibuu}.

\begin{table*}
\centering
\begin{tabular}{|c|c|c|}
\hline
                       & \textbf{G18\_10a}               & \textbf{AR23\_0i}                 \\ \hline
\textbf{Ground State}  & Local Fermi Gas (LFG)           & Custom spectral-function-like LFG \\ \hline
\textbf{Quasi-elastic} & Valencia using dipole expansion & Valencia using z-expansion        \\ \hline
\textbf{2p2h}          & Valencia                        & SuSAv2                            \\ \hline
\textbf{Resonance}     & Berger-Sehgal 2020              & Berger-Sehgal 2020                \\ \hline
\textbf{Inelastic}     & Bodek-Yang 2020                 & Bodek-Yang 2020                   \\ \hline
\textbf{Hadronization} & AGKY 2020                       & AGKY 2020                         \\ \hline
\textbf{FSI}           & hA18                            & hA18                              \\ \hline
\end{tabular}
\caption{The physics models used in the GENIE G18 and AR23 model configurations.}
\label{tab:genie_tunes}
\end{table*}

MicroBooNE observed that G18 underpredicted MiniBooNE's measurement of the cross section for charged-current scattering with zero final-state pions ($\text{CC}0\pi$) taken in the same beam~\cite{ub_tune}. Although the parameters of the nominal GENIE interaction model had already been tuned to make G18, MicroBooNE performed an additional tune of the relevant G18 parameters to a $\text{CC}0\pi$ cross section measurement made in T2K's ND, with very similar flux. The result is an improved prediction of the MiniBooNE data~\cite{ub_tune,t2k_cross_2016}.

\begin{figure*}
\centering
\includegraphics[width=\linewidth]{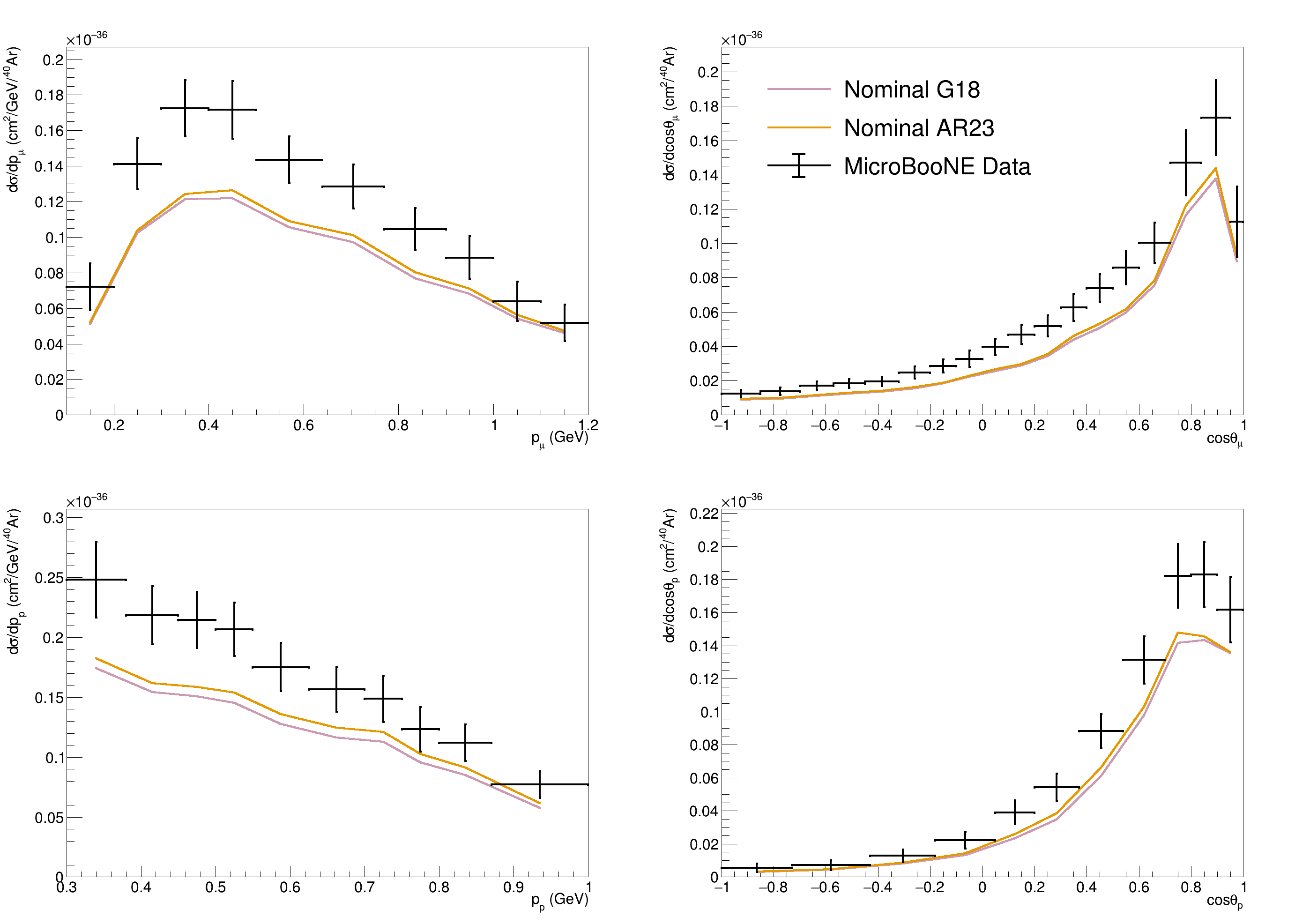}
    \caption{MicroBooNE CC1$\mu$1p single-differential cross section data in muon momentum (top left), muon angle (top right), proton momentum (bottom left), and proton angle (bottom right) plotted with the nominal predictions of the data made by the G18 and AR23 model configurations.}
    \label{fig:nominal_hists}
\end{figure*}

The nominal G18 configuration, which underpredicts MiniBooNE $\text{CC}0\pi$ cross section data, also underpredicts MicroBooNE's measurement of the cross section for charged-current scattering with one final-state proton and zero final-state mesons ($\text{CC}1\mu 1p$). Fig.~\ref{fig:nominal_hists} shows that the nominal AR23 configuration underpredicts the MicroBooNE cross section. We study how either configuration's prediction of the MicroBooNE cross section changes after tuning to cross section measurements made by the T2K, MicroBooNE, and MINERvA experiments~\cite{t2k_cross_2016,t2k_cross_2020,ub_cross,miniboone_xsec}.

The measurement used in the MicroBooNE tune was a T2K 2016 $\text{CC}0\pi$ double-differential cross section in muon kinematic variables~\cite{t2k_cross_2016}. MicroBooNE selected this T2K measurement over measurements made by the ArgoNeuT~\cite{argoneut}, MINERvA~\cite{minerva}, and MiniBooNE~\cite{miniboone_xsec} experiments because ArgoNeuT had low statistics, MINERvA used a higher neutrino energy range, and MiniBooNE used the same neutrino source, which would introduce correlations in the flux between MiniBooNE and MicroBooNE.

The T2K 2016 measurement is reported as a function of muon momentum (p$_\mu$) and the cosine of the muon angle with respect to the direction of the neutrino beam ($\cos\theta_\mu$). In~\cite{t2k_cross_2016}, the cross section is calculated twice using different analysis methods. ``Analysis I'' uses a binned likelihood fit to extract the cross section, while ``Analysis II'' uses  d’Agostini unfolding~\cite{dagostini}. Another difference between the two analysis methods is that Analysis II uses a restricted phase space where p$_{\mu}>0.2$~GeV and cos$\theta_{\mu}>0.6$. MicroBooNE selected the Analysis I cross section for their tune because of its broader phase space.

T2K published an updated CC0$\pi$ cross section in 2020~\cite{t2k_cross_2020}. While the 2016 measurement selected only forward-going muons originating from the upstream fine grained detector (FGD1) that is pure CH, the 2020 measurement selected forward-going, high-angle and backward-going muons originating from either FGD1 or FGD2, giving the 2020 measurement a smaller statistical uncertainty and broader acceptance in muon angle compared to the 2016 measurement. Although FGD2 is composed of alternative layers of CH and H$_2$O, the T2K 2020 cross section used here selects FGD2 neutrino interactions in only the CH layers. We tune G18 and AR23 to the T2K 2016 measurement to check that we can replicate MicroBooNE's result, then to the T2K 2020 measurement to study if anything changes.

We also tune G18 and AR23 to MicroBooNE data. Of course it would not have made sense for MicroBooNE to tune to its own data, nor is it sensible for SBN to tune to data collected in the same beam. However, these MicroBooNE tunes serve as a consistency check for our fitting methods. Specifically, single-differential cross sections for a CC1$\mu$1p signal definition are used, four of which are plotted in Fig.~\ref{fig:nominal_hists}. The cross sections as a function of $p_\mu$ and $\theta_\mu$ are fit because they allow for the closest comparison to the double-differential T2K measurements, which are also reported as a function of lepton kinematics. These two results are not independent because they use the same signal selection. In particular the normalization of the two MicroBooNE measurements is expected to be fully correlated.

Finally, the two GENIE configurations are tuned to a MINERvA CC0$\pi$ measurement of the muon neutrino cross section on hydrocarbon from the ``low-energy'' tune of the NUMI beam (Fermilab's second neutrino beam) with a peak flux of 3.5~GeV~\cite{minerva}. This double-differential cross section is reported as a function of the longitudinal and transverse muon momentum ($p_{||}$ and $p_T$, respectively). Table~\ref{tab:xsec} summaries the five $\nu_\mu$ cross section measurements used in this study.

The MINERvA data are at notably higher neutrino energy compared to T2K and MicroBooNE, providing a test of the neutrino energy dependence of the cross section models and the robustness of the uncertainty parameterizations at simultaneously describing data at different energies. The higher-energy data also has a different mix of underlying interaction processes, and is predicted to have a larger contribution from pion production where the pion is absorbed.

\begin{table*}[]
\centering
\begin{tabular}{|c|c|c|c|c|c|}
\hline
                    & \textbf{Beam} & \textbf{Target} & \textbf{Signal} & \textbf{$\sigma$-dimensions} & \textbf{$\sigma$-variables} \\ \hline
\textbf{T2K 2016~\cite{t2k_cross_2016}}   & $\nu_\mu$     & CH              & CC$0\pi$        & 2D                           & $(\cos \theta_\mu ,p_\mu )$ \\ \hline
\textbf{T2K 2020~\cite{t2k_cross_2020}}   & $\nu_\mu$     & CH              & CC$0\pi$        & 2D                           & $(\cos \theta_\mu ,p_\mu )$ \\ \hline
\textbf{MicroBooNE~\cite{ub_cross}} & $\nu_\mu$     & Ar              & CC$1\mu 1p$     & 1D                           & $p_\mu$, $\cos \theta_\mu$  \\ \hline
\textbf{MINERvA~\cite{minerva}}    & $\nu_\mu$     & CH              & CC$0\pi$        & 2D                           & $(p_{\mu ,L},p_{\mu ,T})$   \\ \hline
\textbf{MiniBooNE~\cite{miniboone_xsec}}  & $\nu_\mu$     & CH$_2$              & CC$0\pi$        & 2D                           & $(\cos \theta_\mu ,T_\mu )$ \\ \hline
\end{tabular}
\caption{A summary of the cross section measurements used in this study.}
\label{tab:xsec}
\end{table*}

For G18, the fit parameters used in this study are the same as those used by MicroBooNE because they have the greatest impact on the $\text{CC}0\pi$ cross section~\cite{ub_tune}: MaCCQE, NormCCMEC, RPA\_CCQE, and XSecShape\_CCMEC. The MaCCQE parameter adjusts the value of the axial mass in the dipole parameterization of the CCQE axial-vector form factor. The NormCCMEC parameter adjusts the normalization of the cross section of charged-current (CC) 2p2h interactions, which are also referred to as meson exchange currents (MEC). The RPA\_CCQE parameter adjusts the strength of the suppression of the CCQE cross section at low Q$^2$ due to nucleon-nucleon long range correlations. The XSecShape\_CCMEC parameter adjusts the shape of the 2p2h cross section between the Valencia model~\cite{valencia} and the Empirical model~\cite{empirical}. 

Due to the differences between G18 and AR23 summarized in Table~\ref{tab:genie_tunes}, the fit parameters in AR23 are different. Six parameters explain nearly the entire variation in the CC$0\pi$ cross section at these energies: RPA\_CCQE, NormCCMEC, ZExp\_PCA\_b1, b2, b3, and b4. These RPA\_CCQE and NormCCMEC parameters have the same function in AR23 as in G18. The ZExp\_PCA\_b1, b2, b3, and b4 parameters adjust the coefficients in the z-expansion parameterization of the CCQE axial-vector form factor. Although the four coefficients are necessarily correlated, the corresponding fit parameters have been transferred to an uncorrelated basis by means of principal component analysis (PCA).

The fitting for this study is performed with the NUISANCE software package~\cite{nuisance}, which uses a gradient descent algorithm from the MINUIT package~\cite{minuit} to minimize the $\chi^2$ test-statistic with respect to the chosen parameters. The $\chi^2$ is calculated in two different ways. The ``full $\chi^2$'', which uses the covariance matrix from the data and adds a penalty term for pulls on the model parameters, is given by Eqn.~\ref{eq:chi2full}:

\begin{equation}
    \chi^2_{full}=\sum_{i}^{N_d}\sum_{j}^{N_d}(D_i-P_i)(\textbf{C}^{-1})_{ij}(D_j-P_j) + \sum_{i}^{N_p}\sigma_{i}^{2}
    \label{eq:chi2full}
\end{equation}

\noindent
where $N_d$ is the number of data points, $D_i$ is the $i$th data point, $P_i$ is the prediction of the $i$th data point, \textbf{C} is the covariance matrix describing the data uncertainty such that $(\textbf{C}^{-1})_{ij}$ is the $ij$th element of its inverse, $N_p$ is the number of fit parameters, and $\sigma_{i}$ is the $i$th best-fit parameter value, expressed in units of 1 sigma uncertainty. 

An alternative test statistic is the ``diagonal $\chi^2$'' used by MicroBooNE~\cite{ub_tune}, which uses only the diagonal elements of the covariance matrix and does not add a penalty term like in $\chi^2_{full}$:
\begin{equation}
    \chi^2_{diag}=\sum_{i}^{N_d}\frac{(D_i-P_i)^2}{C_{ii}}.
    \label{eq:chi2diag}
\end{equation}

Although the standard practice in neutrino physics is to use $\chi^2_{full}$ as the test statistic, MicroBooNE opted to use $\chi^2_{diag}$ when they tuned G18, a fitting method that will henceforth be referred to as the ``diagonal G18'' method. MicroBooNE explained their non-standard choice as the solution to the problem that fitting T2K 2016 with the full covariance matrix resulted in a significant reduction in the total cross section prediction or Peelle's Pertinent Puzzle (PPP)~\cite{ppp}. PPP states that the weighted mean of experimental data with systematic uncertainties may fall outside the range of the data. In MicroBooNE's case, PPP caused a fit that visibly agreed with the data to be disfavored compared to a fit with a lower cross section that visibly disagrees with the data but has a lower test statistic due to strong correlations in the data uncertainties.

MicroBooNE found that disregarding the off-diagonal elements of the covariance matrix associated with the T2K 2016 cross section avoided PPP~\cite{ub_tune}. In this work, we replicate MicroBooNE's diagonal G18 tune and compare the results to those obtained using the full covariance matrix or a ``full G18'' tune. We also look at the results of tuning AR23 using the full covariance matrix (referred to as the ``full AR23'' method). For consistency with MicroBooNE,  the penalty term from Eqn.~\ref{eq:chi2full} is not included in our diagonal G18 method. However, we do use a penalty term in the full G18 and full AR23 methods because it is the conventional way to prevent fit parameters from running far away from their nominal values.

\section{Results and Discussion}\label{sec:results}

\begin{figure}
\centering
\includegraphics[width=\linewidth]{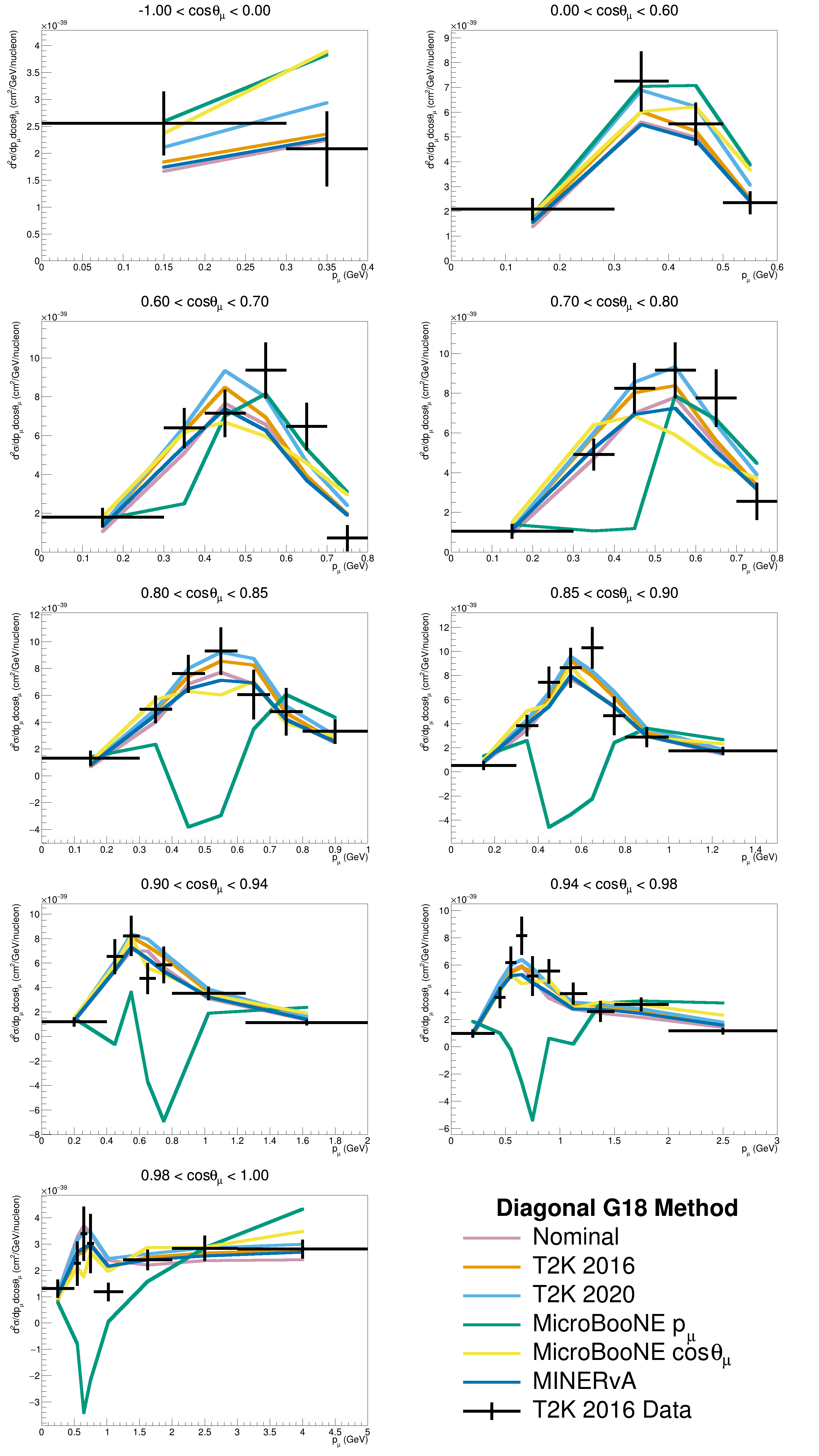}
    \caption{The T2K 2016 data is compared to predictions from the nominal G18 model configuration, as well as five different tunes obtained by fitting G18 to the data given in the legend using the ``diagonal'' method.}
    \label{fig:t2k_2016_i_diag_g18}
\end{figure}

Using the diagonal method, G18 is fit to T2K 2016 I, T2K 2020, MINERvA, MicroBooNE p$_\mu$, and MicroBooNE $\cos\theta_\mu$. Fig.~\ref{fig:t2k_2016_i_diag_g18} shows how these five tunes perform compared to the nominal G18 model when predicting the T2K 2016 data. The orange curve in the figure is the result of tuning the G18 model to the data shown. The other curves come from applying the tuning method of Ref.~\cite{ub_tune} to other measurements, including MicroBooNE's own data, then predicting T2K 2016 using the tuned parameters. The green curve in Fig.~\ref{fig:t2k_2016_i_diag_g18} is essentially the inverse of Ref.~\cite{ub_tune}: the model is tuned to MicroBooNE data and the result is then compared to T2K. This yields an unphysical result; the tuned parameter values from the MicroBooNE $p_\mu$ fit predict unphysical, negative cross sections in several regions of the T2K parameter space. This is the result of a poor fit to the MicroBooNE $p_\mu$ measurement using the diagonal G18 method.

\begin{figure*}
\centering
\includegraphics[width=\linewidth]{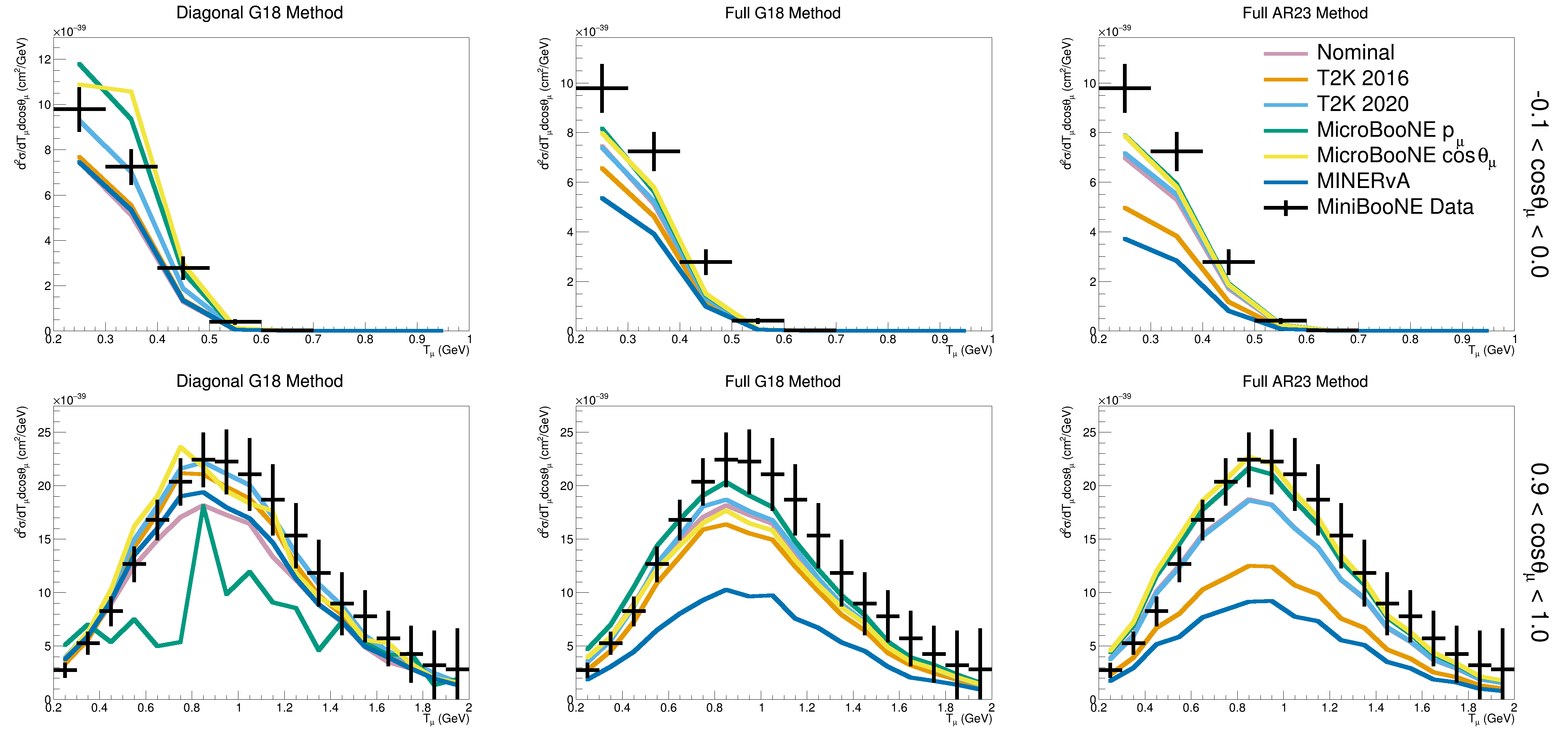}
    \caption{MiniBooNE data in the $-0.1<\cos\theta_\mu<0.0$ bin (top) and $0.9<\cos\theta_\mu<1.0$ bin (bottom) is compared to predictions from the nominal G18 and AR23 models and their tunes to the data in the legend using the method in the plot title}
    \label{fig:miniboone_xsec}
\end{figure*}

MicroBooNE tuned G18 to T2K 2016 in order to improve its prediction of the MiniBooNE CCQE-like ($\text{CC}0\pi$) measurement of the muon neutrino cross section on hydrocarbon~\cite{miniboone_xsec}. Fig.~\ref{fig:miniboone_xsec} shows the nominal predictions as well as the results of tuning G18 to each of the five measurements using the diagonal G18 method, and the full covariance with both G18 and AR23; two bins of muon angle are shown, one corresponding to forward angles and the other to wide angles, as a representation of the full 2D dataset. Diagonal G18 reproduces the MicroBooNE result in Ref.~\cite{ub_tune}, namely that tuning to T2K 2016 increases the cross section and describes the MiniBooNE data well. However, applying the same method to the MicroBooNE $p_\mu$ data gives an unphysical result. Including the full covariance matrix, the G18 tunes do a visibly worse job of predicting the MiniBooNE data. 

\begin{figure}
\centering
\includegraphics[width=\linewidth]{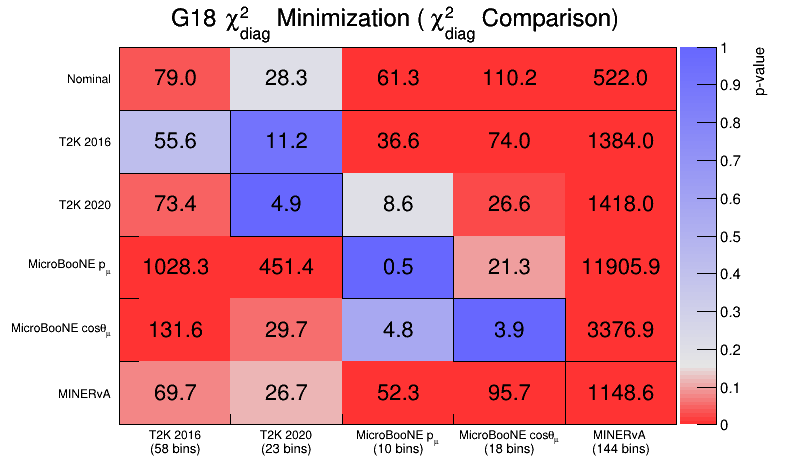}
    \caption{The $\chi^2_{diag}$ from Eqn.~\ref{eq:chi2diag} obtained by comparing the G18 model configuration to five different measurements, indicated by each column. The top row is the nominal prediction, subsequent rows are the best-fit tune to the measurement indicated by the row, using the diagonal method. The cell color is a $p$-value assuming $\chi^2_{diag}$ follows a $\chi^2$ distribution.}
    \label{fig:nuiscomp_pvalue_double_diag_g18}
\end{figure}

Fig.~\ref{fig:nuiscomp_pvalue_double_diag_g18} shows how the five diagonal G18 tunes perform when predicting all five data sets. The top row holds the nominal, un-tuned predictions, while each subsequent row holds the predictions of a different diagonal G18 tune, and each column contains a different cross section measurement. Printed in each cell is the $\chi^{2}_{diag}$ of Eqn.~\ref{eq:chi2diag} for the prediction indicated by the column using the tune indicated by the row. Five of the cells show the actual minimum $\chi^{2}_{diag}$ calculated during the tuning process, while the other cells show the goodness of fit of a prediction to one measurement after a fit to another measurement. The two MicroBooNE single-differential cross sections share a common signal sample and thus are not independent, so tuning to one and predicting another falls somewhere in between. The normalizations of these two measurements are correlated, and there are also correlations between the muon energy and angle.

The number of degrees of freedom (ndof) in $\chi^{2}_{diag}$ varies significantly between columns because the number of bins (printed below the column label) varies significantly between cross section measurements. As a visual aid in interpreting the level of agreement, a $p$-value is computed assuming that $\chi^2_{diag}$ follows a $\chi^2$ distribution. For nominal predictions, the number of degrees of freedom is equal to the number of bins in the data, while the number of bins minus four is used for tuned predictions in order to account for the four free parameters in the G18 tuning procedure. The cells of Fig.~\ref{fig:nuiscomp_pvalue_double_diag_g18} are colored according to this $p$-value such that blueish cells represent reasonable agreement and reddish cells represent poor agreement between the data and the nominal or tuned model. A high $p$-value indicates that the tuned model is able to describe the data accurately, but not necessarily that the free parameters are close to their nominal values. 

\begin{figure}
\centering
\includegraphics[width=\linewidth]{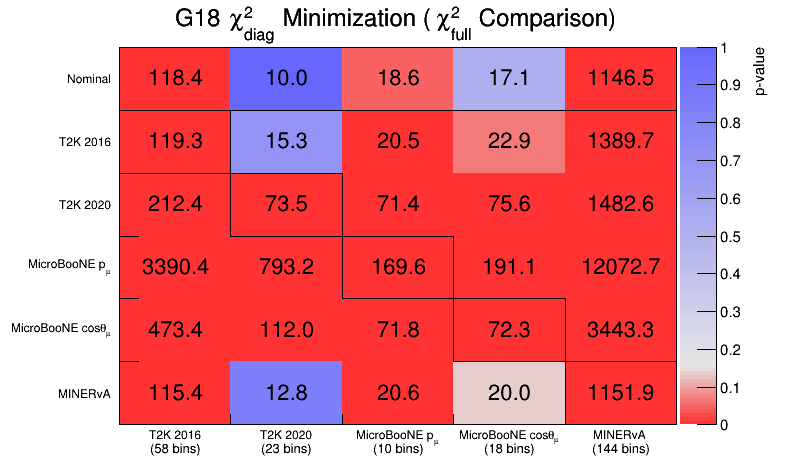}
    \caption{The $\chi^2_{full}$ from Eqn.~\ref{eq:chi2full} obtained by comparing the G18 model configuration to five different measurements, indicated by each column. The top row is the nominal prediction, subsequent rows are the best-fit tune to the measurement indicated by the row, using the diagonal method. The cell color is a $p$-value assuming $\chi^2_{full}$ follows a $\chi^2$ distribution.}
    \label{fig:nuiscomp_pvalue_diag_g18}
\end{figure}

Fig.~\ref{fig:nuiscomp_pvalue_diag_g18} has the same format as Fig.~\ref{fig:nuiscomp_pvalue_double_diag_g18} and uses the same diagonal G18 tunes, but shows the $\chi^{2}_{full}$ of Eqn.~\ref{eq:chi2full}, which uses the entire covariance matrix and adds a penalty term for pulls on the model parameters. In this case where goodness of fit is assessed using $\chi^{2}_{full}$, none of the diagonal G18 tuned models are an improvement over the nominal model. This suggests that the range of cross section shapes allowed by the systematic uncertainties on the data is not represented well by any diagonal G18 tune, and that the best fit obtained by minimizing $\chi^2_{diag}$ does not actually describe the data very well when the correlations in the data systematics are taken into account.

\begin{figure}
\centering
\includegraphics[width=\linewidth]{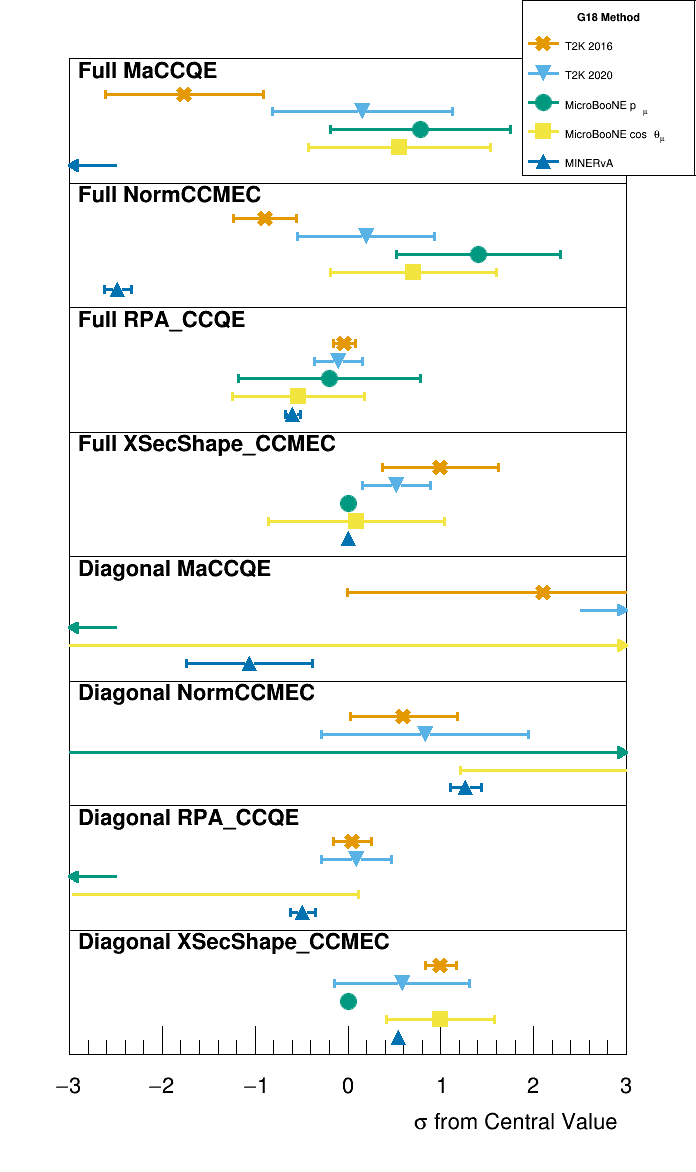}
\caption{Post-fit parameter values and errors (in units of 1$\sigma$ uncertainty) obtained from the full G18 and diagonal G18 fitting methods and the indicated cross section measurements.}
\label{fig:par_val_g18}
\end{figure}

The parameter pulls used in each of the rows of Figs.~\ref{fig:nuiscomp_pvalue_double_diag_g18} and~\ref{fig:nuiscomp_pvalue_diag_g18} are plotted in Fig.~\ref{fig:par_val_g18} along with their post-fit uncertainties. The unit of the parameter pull is the \textit{a priori} 1$\sigma$ uncertainty. The diagonal G18 tune to T2K 2016 produces reasonable results, as was observed by MicroBooNE. However, the tunes to the other four measurements result in poor fits with pull values greater than three standard deviations or unreasonably small pull errors. In particular, the tune to MicroBooNE $p_\mu$ pulls three of the four parameters outside of the prior $\pm 3\sigma$ range, causing the unphysical results in Fig.~\ref{fig:t2k_2016_i_diag_g18}.

\begin{figure}
\centering
\includegraphics[width=\linewidth]{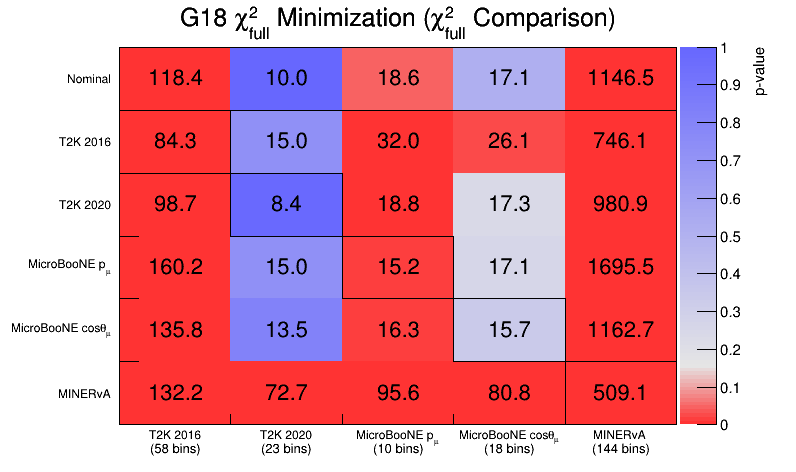}
    \caption{The $\chi^2_{full}$ from Eqn.~\ref{eq:chi2full} obtained by comparing the G18 model configuration to five different measurements, indicated by each column. The top row is the nominal prediction, subsequent rows are the best-fit tune to the measurement indicated by the row, using the full covariance matrix of the data. The cell color is a $p$-value assuming $\chi^2_{full}$ follows a $\chi^2$ distribution.}
    \label{fig:nuiscomp_pvalue_full_g18}
\end{figure}

Fig.~\ref{fig:par_val_g18} also shows the parameter pulls used in the five full G18 tunes. With the data covariance matrix included, the tunes to T2K 2020 and both MicroBooNE measurements produce reasonable results, while the tunes to T2K 2016 and MINERvA result in poor fits. As before, each set of full G18 parameter pulls is used to predict the five cross section measurements, and the $p$-values of these predictions are plotted in Fig.~\ref{fig:nuiscomp_pvalue_full_g18}. The fit to T2K 2016 is notably poor, while the model describes T2K 2020 and both MicroBooNE measurements reasonably well out-of-the-box, and especially with parameters tuned to any of the data except MINERvA.

\begin{figure}
\centering
\includegraphics[width=\linewidth]{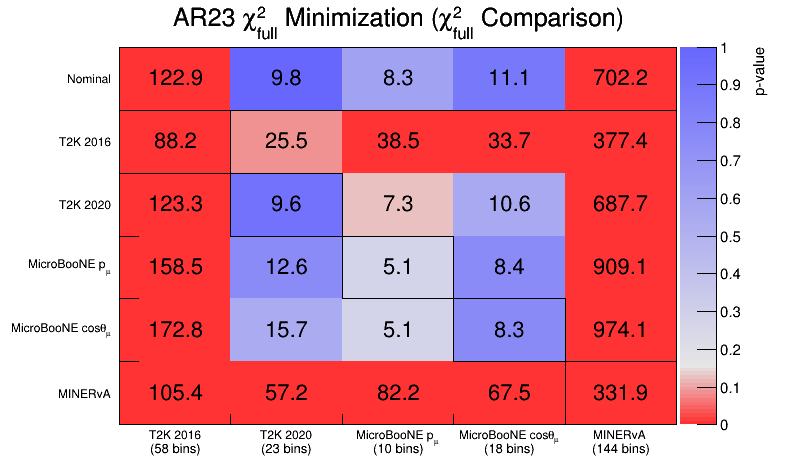}
    \caption{The $\chi^2_{full}$ from Eqn.~\ref{eq:chi2full} obtained by comparing the AR23 model configuration to five different measurements, indicated by each column. The top row is the nominal prediction, subsequent rows are the best-fit tune to the measurement indicated by the row, using the diagonal method. The cell color is a $p$-value assuming $\chi^2_{full}$ follows a $\chi^2$ distribution.}
    \label{fig:nuiscomp_pvalue_full_AR23}
\end{figure}

\begin{figure}
\centering
\includegraphics[width=\linewidth]{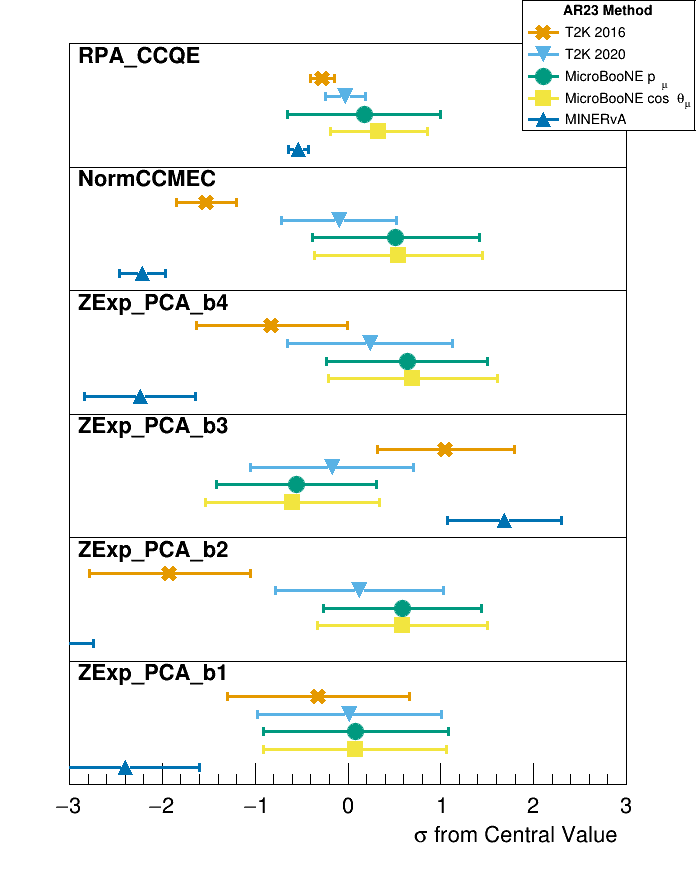}
\caption{Post-fit parameter values and errors (in units of 1$\sigma$ uncertainty) obtained by fitting AR23 using the full covariance matrix to the indicated cross section measurements.}
\label{fig:par_val_full_AR23}
\end{figure}

Fig.~\ref{fig:par_val_full_AR23} shows the parameter pulls for fits using AR23, and Fig.~\ref{fig:nuiscomp_pvalue_full_AR23} gives the fit quality. The AR23 configuration does slightly better in describing the MicroBooNE momentum spectrum compared to G18. It is in good agreement with T2K 2020 but not T2K 2016, and fits to the 2016 data yield a poor prediction of every other measurement. The parameter values for the fits to T2K 2020 and MicroBooNE are broadly consistent, indicating that AR23 is able to simultaneously describe lepton kinematics in this energy regime on two different nuclei.

\begin{figure*}
\centering
\includegraphics[width=\linewidth]{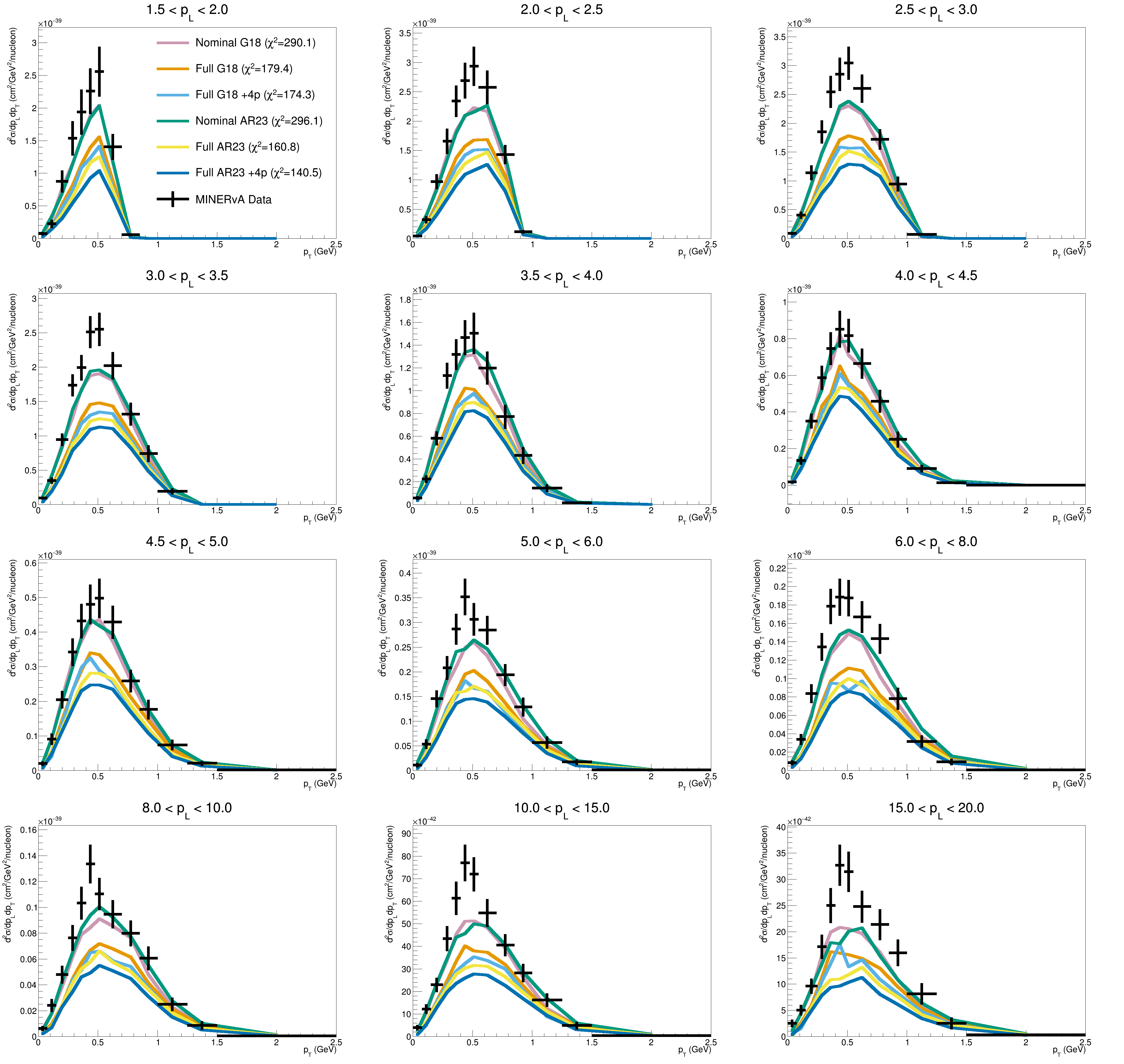}
    \caption{MINERvA data is compared to predictions from the G18 and AR23 models, showing the nominal predictions, the result of tuning to the data with the full covariance matrix, and the tune with two additional pion production parameters.}
    \label{fig:mnv_comp}
\end{figure*}

Neither model configuration is able to describe the MINERvA data, which can be seen by the high $\chi^2$ in the bottom row of Figs.~\ref{fig:nuiscomp_pvalue_full_g18} and \ref{fig:nuiscomp_pvalue_full_AR23}. Fig.~\ref{fig:mnv_comp} plots the MINERvA data with the nominal G18 and AR23 predictions as $p_T$ distributions in bins of $p_L$. Because the MINERvA neutrino beam is peaked around 3.5 GeV, most of the cross section falls in the first few bins of $p_L$. In this region, there is disagreement between data and MC at moderate $p_T$.

Tuning either model configuration to the MINERvA data decreases the total cross section, giving an apparently worse fit, albeit with a smaller $\chi^2$. This is qualitatively similar to what MicroBooNE observed in fitting T2K 2016 and attributed to Peelle's Pertinent Puzzle (PPP)~\cite{ub_tune}. The PPP visible in Fig.~\ref{fig:mnv_comp} is a sign that the model parameters used in the low-energy T2K 2020 and MicroBooNE tunes do not have sufficient freedom to describe the high-energy MINERvA cross section and its (strongly correlated) systematic uncertainties.

Compared to T2K 2020 and MicroBooNE, the MINERvA cross section has a different mix of interaction processes with a much higher contribution from pion production followed by absorption. It might be expected that additional parameters are required in order to model these contributions. As a test, fits are performed with four additional parameters that govern pion production: the vector and axial masses MvCCRES and MaCCRES, the fraction of pions inelastically scattered FrInel\_pi, and the fraction of pions absorbed FrAbs\_pi. However, the extra freedom does not improve the result in Fig.~\ref{fig:mnv_comp}.

\section{Tuning in LBL and SBL experiments}\label{sec:tuningLBLSBL}

The neutrino flux appears in the denominator of the formula used to extract cross sections. For a differential measurement with respect to some quantity $X$, which is unfolded to ``true'' quantities, the cross section in a bin $i$ can be written as

\begin{equation}\label{eqn:cross_section}
    \frac{d\sigma}{dX_i}=\frac{\sum_j U_{ij}(N_j-N^{bkg}_j)}{\epsilon_i T\Phi \Delta X_i},
\end{equation}
where $N_j$ is the number of selected events in a reconstructed bin $j$, $N^{bkg}_j$ is the predicted background in that bin, $U_{ij}$ is an element of the unfolding matrix that relates true to reconstructed $X$, $\epsilon_i$ is the selection efficiency, $T$ is the number of targets, $\Phi$ is the neutrino flux, and $\Delta X_i$ is the bin width. 

In all cross section data ever reported, the flux $\Phi$ in the denominator of Eqn.~\ref{eqn:cross_section} is an estimate of the un-oscillated flux of the neutrino source assuming the probability of short-baseline oscillations is zero. For an LBL experiment, tuning the interaction model to a flux-integrated SBL cross section is consistent with an oscillation analysis that neglects short-baseline oscillations. The ND data is then used to constrain the flux and cross section uncertainties with the same assumption, namely that oscillations are not present in the ND.

For an SBL experiment, however, tuning the interaction model to external data is problematic, because available cross section data is analyzed assuming there are no short-baseline oscillations. If a sterile oscillation were present in an SBL experiment, then any cross section measurement with an overlap in $L/E_\nu$ (see Table~\ref{tab:experiments}) would be dividing by the incorrect flux due to that oscillation. 

Consider a hypothetical where an experiment has a perfect interaction model, and 10\% of active $\nu_{\mu}$ oscillate to sterile neutrinos with sufficiently high $\Delta m^2$ that the oscillations are extremely rapid and only average effects are observed for $L/E_\nu$ of order 0.1-10 km/GeV. If a perfect interaction model is tuned to external data, the flux assumed in that external measurement will be 10\% higher than the actual active $\nu_{\mu}$ flux at the detector, so the extracted cross section will be 10\% too low. This discrepancy would be attributed to cross section effects, and tuned away. The tuned model will then agree perfectly with the SBL data and the experiment will publish an exclusion on SBL oscillations. 

Alternatively, an SBL experiment could tune its model to external data from a different range of $L/E_\nu$ than that to which the SBL experiment is sensitive. This is limited by the availability of data, and the typically broad range of $L/E_\nu$ that the data span. In this scenario, the SBL experiment should only publish limits in the range of $\Delta m^2$ where the effect of oscillations on the cross section data is negligible.

A more subtle risk is the use of pre-tuned interaction models in SBL searches. GENIE, for example, has numerous built-in comprehensive model configuration tunes~\cite{genie_tune}. These tunes are extremely convenient for LBL experiments like NOvA~\cite{nova_genie_tune}. However, the data used in these tunes is analyzed with the assumption of no sterile oscillations, so their use in SBL searches could potentially mask the sterile oscillation signal in the same way as above.

In the specific case of SBN, the two-detector SBL search at Fermilab, any tuning of the interaction model to external SBL cross section data would be an unnecessary harm. Unlike its predecessors (all one-detector SBL searches), SBN will be able to resolve discrepancies between its interaction model and its data internally utilizing its near detector SBND, which sits at a different range of $L/E_\nu$ than its far detector ICARUS, in an analysis that will explicitly control for sterile oscillations in SBND's data.

The best course of action for SBN is to choose a base model that is as free as possible from the influence of external SBL data. Instead of tuning that model, which would in principle improve its agreement with data, SBN should check that there is sufficient freedom in the model's uncertainties to describe all of the available external data, for example by following the methods presented in this work. The consequence of not tuning may be large \textit{a priori} cross section uncertainties, but these can be constrained with data from SBND. Unlike external cross section data, the SBND constraint can be implemented in a joint fit with ICARUS data, such that the presence of SBL oscillations can be handled correctly.

Although this section has focused on the risks of tuning in SBL experiments, LBL experiments should also beware. The observation that tunes to T2K 2020 yield excellent descriptions of that data set, but poor descriptions of the MINERvA data set, which has the same target nucleus and nearly the same signal definition, should caution DUNE against over-relying on cross section data from SBND. While precise measurements from SBND will obviously be valuable for informing model selection for DUNE analyses, it remains critical for DUNE to rely primarily on measurement from its own near detector.

\section{Conclusions}\label{sec:conclusion}

\begin{itemize}
    \item The T2K 2020 and MicroBooNE data are mutually well described by the AR23 configuration and its uncertainties. The G18 configuration also does a reasonable job of simultaneously describing T2K 2020 and MicroBooNE, but with some tension in the MicroBooNE muon momentum.
    \item The T2K 2016 measurement is inconsistent with the others, including the T2K 2020 measurement that uses some of the same data.
    \item Neither G18 nor AR23 are able to describe the MINERvA data. This suggests that these models are adequate at low energy but fall short at higher energy, or fail to model the additional processes that are present in MINERvA but not at lower energies.
    \item The result of the MicroBooNE tuning procedure is strongly dependent on the specific data chosen, and would be substantially different if the more recent T2K 2020 data were used. The procedure yields unrealistic results when applied to other measurements, including MicroBooNE's own data.
    \item Short-baseline oscillation experiments should be extremely careful about tuning to cross section data. The two-detector SBN should \textit{not} tune to data, and should use a model configuration that is not implicitly tuned to data, like AR23.
\end{itemize}

\section{Acknowledgments}

We thank Jaesung Kim for assistance with installing software. We thank Jaesung Kim and Faiza Akbar for feedback on a draft of this manuscript.

This material is based on work supported by the U.S. Department of Energy, Office of Science, Office of High Energy Physics under Award Numbers DE-SC0008475 and DE-SC0024431; by the U.S. Department of Energy, Office of Science, Office of Workforce Development for Teachers and Scientists, Office of Science Graduate Student Research (SCGSR) program; and by the University of Rochester. The SCGSR program is administered by the Oak Ridge Institute for Science and Education for the DOE under contract number DE‐SC0014664.

\section{Disclaimer}

This report was prepared as an account of work sponsored by an agency of the United States Government.  Neither the United States Government nor any agency thereof, nor any of their employees, makes any warranty, express or implied, or assumes any legal liability or responsibility for the accuracy, completeness, or usefulness of any information, apparatus, product, or process disclosed, or represents that its use would not infringe privately owned rights.  Reference herein to any specific commercial product, process, or service by trade name, trademark, manufacturer, or otherwise does not necessarily constitute or imply its endorsement, recommendation, or favoring by the United States Government or any agency thereof.  The views and opinions of authors expressed herein do not necessarily state or reflect those of the United States Government or any agency thereof.

\clearpage
\bibliography{main}

\clearpage
\appendix
\end{document}